\documentclass[twocolumn,showpacs,preprintnumbers,prb,fleqn,superscriptaddress]{revtex4}

\usepackage{graphicx}
\usepackage{dcolumn}
\usepackage{amsmath}
\usepackage{amsfonts}
\usepackage{bm}

\newcommand*{\win}{\omega_\mathrm{in}}
\newcommand*{\omph}{\omega_\mathrm{ph}}
\newcommand*{\Ai}{\mathop{\mathrm{Ai}}\nolimits}
\newcommand*{\Bi}{\mathop{\mathrm{Bi}}\nolimits}
\renewcommand{\vec}[1]{\mathbf{#1}}
\newcommand{\sign}{\mathop{\mathrm{sgn}}}

\newcommand{\ep}{\epsilon}
\newcommand{\Tr}{\mathop{\mathrm{Tr}}}
\newcommand{\unitmatrix}{\openone}

\begin{document}

\title{Effect of a magnetic field on the two-phonon Raman scattering in graphene}

\author{C. Faugeras}
\affiliation{LNCMI-CNRS, BP 166, 38042 Grenoble, France}

\author{P. Kossacki}
\affiliation{LNCMI-CNRS, BP 166, 38042 Grenoble, France}
\affiliation{Institute of Experimental Physics, University of Warsaw, Poland}

\author{D. M. Basko}
\affiliation{Laboratoire de Physique et Mod\'elisation des Milieux Condens\'es, Universit\'e Joseph Fourier and CNRS,
25 rue des Martyrs, 38042 Grenoble, France}

\author{M. Amado}
\affiliation{LNCMI-CNRS, BP 166, 38042 Grenoble, France}
\affiliation{QNS-GISC, Departamento de F\'{i}sica de Materiales, Universidad Complutense,
E-28040 Madrid, Spain}

\author{M. Sprinkle}
\affiliation{School of Physics, Georgia Institute of Technology, Atlanta, Georgia 30332, USA}

\author{C. Berger}
\affiliation{School of Physics, Georgia Institute of Technology,
Atlanta, Georgia 30332, USA}
\affiliation{CNRS-Institut N$\acute{e}$el, BP 166, 38042 Grenoble Cedex 9,
France.}

\author{W.A. de Heer}
\affiliation{School of Physics, Georgia Institute of Technology, Atlanta, Georgia 30332, USA}

\author{M. Potemski}
\affiliation{LNCMI-CNRS, BP 166, 38042 Grenoble, France}

\begin{abstract}
We have studied, both experimentally and theoretically, the change
of the so-called $2D$ band of the Raman scattering spectrum of
graphene (the two-phonon peak near 2700~cm$^{-1}$) in an external
magnetic field applied perpendicular to the graphene crystal
plane at liquid helium temperature. A shift to lower frequency and broadening of this band is
observed as the magnetic field is increased from 0 to 33~T. At
fields up to 5--10~T the changes are quadratic in the field while
they become linear at higher magnetic fields. This effect is
explained by the curving of the quasiclassical trajectories of the
photo-excited electrons and holes in the magnetic field, which
enables us (i)~to extract the electron inelastic scattering rate,
and (ii)~to conclude that electronic scattering accounts for about
half of the measured width of the $2D$~peak.
\end{abstract}

 \maketitle

\section{Introduction}

Graphene, a two-dimensional plane of carbon atoms arranged in a
honeycomb lattice, has stimulated many experimental and
theoretical efforts to study and understand its exotic electronic
and optical properties~\cite{Geim2007}. Graphene can be produced
by mechanical exfoliation of bulk graphite~\cite{Novoselov2005},
by the decomposition of a SiC substrate at high
temperatures,\cite{Berger2004} or by epitaxial growth on different
metal surfaces~\cite{Sutter2008}. Graphene on SiC is well adapted
for optical measurements as samples are of macroscopic size and
lie on a substrate which can be insulating, and hence, transparent
in a wide, from far-infrared to visible spectral range.

Raman scattering is a key element to graphene studies as it is a
reliable and non-destructive technique to establish the monolayer
character of a graphene specimen~\cite{Ferrari2006,Graf2006} or
the negligeable interlayer coupling in multilayer epitaxial
graphene (MEG) samples \cite{Faugeras2008}. This relies on the
analysis of the  shape of the observed $2D$ band feature (the
two-phonon peak around 2700~cm$^{-1}$, also referred as $D^*$ or
$G'$) which is the second-order overtone of the $D$~band.
While defects (in a general meaning -- impurities, interface
roughness, etc.) are needed for the $D$ band to be observed, the
$2D$ band is always seen and it even appears as the most
pronounced feature in the Raman scattering spectrum of graphene.
Because of the doubly-resonant nature of the
$D$~band\cite{Baranov1987,Thomsen2000} and the fully-resonant
nature of the $2D$ band,\cite{Kurti2002,Basko2007} they both show
a dispersive behavior with the excitation energy, what allowed to
tracing the phonon band structure of different carbon
allotropes\cite{Saito2002,Mafra2009}. In this paper, we show that
a strong magnetic field applied perpendicular to the plane of a
graphene crystal significantly affects the $2D$ band feature.
Increasing the magnetic field up to 33 T, we observe a
simultaneous red shift of the $2D$ band energy and a strong
broadening of the peak.

In the following, we show that the evolution of the $2D$ band in
magnetic field can be understood and modelled quantitatively as a
pure orbital effect of the magnetic field on the intermediate
states of the Raman process, i.~e., electrons and holes. The field
induces circular orbits to the photo-generated electron-hole
pairs, and this modifies the momenta of the optical phonons
emitted during the Raman scattering process. This leads to a red
shift and to a broadening of the $2D$~band feature as the magnetic
field is increased, an effect that we observe experimentally and
describe theoretically below. We note that this effect is not
specific to a graphene monolayer as the two essential ingredients
are the fully-resonant nature of the $2D$~peak, and the free
motion of electrons in the plane. Thus, it should also be
characteristic for the multilayer graphene and graphite.

\section{Samples and experiment}

Raman scattering measurements were performed in the backscattering
geometry at liquid helium temperature and in magnetic fields up to
33~T applied perpendicular to the plane of the graphene crystal. A
Ti:Saphire laser tuned at an accurately controlled wavelength of
720~nm (corresponding to the incident photon energy
$\hbar\win=1.722\:\mbox{eV}$) was used for excitation. Optical
fibers, both with a core diameter of 200 $\mu$m, were used for
excitation and for collection. The resulting laser spot on the
sample had a diameter of $\sim$600 $\mu$m with a typical power of
$\sim$100 mW.

The samples presented in this study are MEG grown by the thermal
decomposition of the carbon face of a SiC substrate
\cite{Berger2004}. Despite the fact that these samples contain
$\sim$70 layers, simple Dirac-like electronic bands, such as the
ones found in exfoliated graphene mono-layers, persist in these
highly graphitized structures because of the particular rotational
stacking exhibited by the adjacent graphitic
planes~\cite{Hass2008}. Dirac Fermions in such MEG structures have
been evidenced by magneto-transport~\cite{Berger2006}, by
magneto-transmission~\cite{Sadowski2006} and, more recently, by
scanning tunnelling spectroscopy~\cite{Miller2009} and by
magneto-Raman scattering experiments~\cite{Faugeras2009}. Apart
from the first few interface layers which are highly doped due to
the charge transfer from the SiC substrate~\cite{Berger2006}, the
majority layers, which we probe with Raman scattering, are
quasi-neutral with densities as low as $5\times10^9$ cm$^{-2}$ and
mobilities as high as
$250000\:\mbox{cm}^2(\mbox{V}\cdot\mbox{s})^{-1}$, as deduced from
magneto-transmission experiments~\cite{Orlita2008}. MEG samples,
and especially the highly graphitized specimens, are not
homogenous on the scale of our laser spot and contain some
Bernal-stacked inclusions as observed through micro-Raman
scattering experiments~\cite{Faugeras2008}.

\begin{figure}
\includegraphics[width=1\linewidth,angle=0,clip]{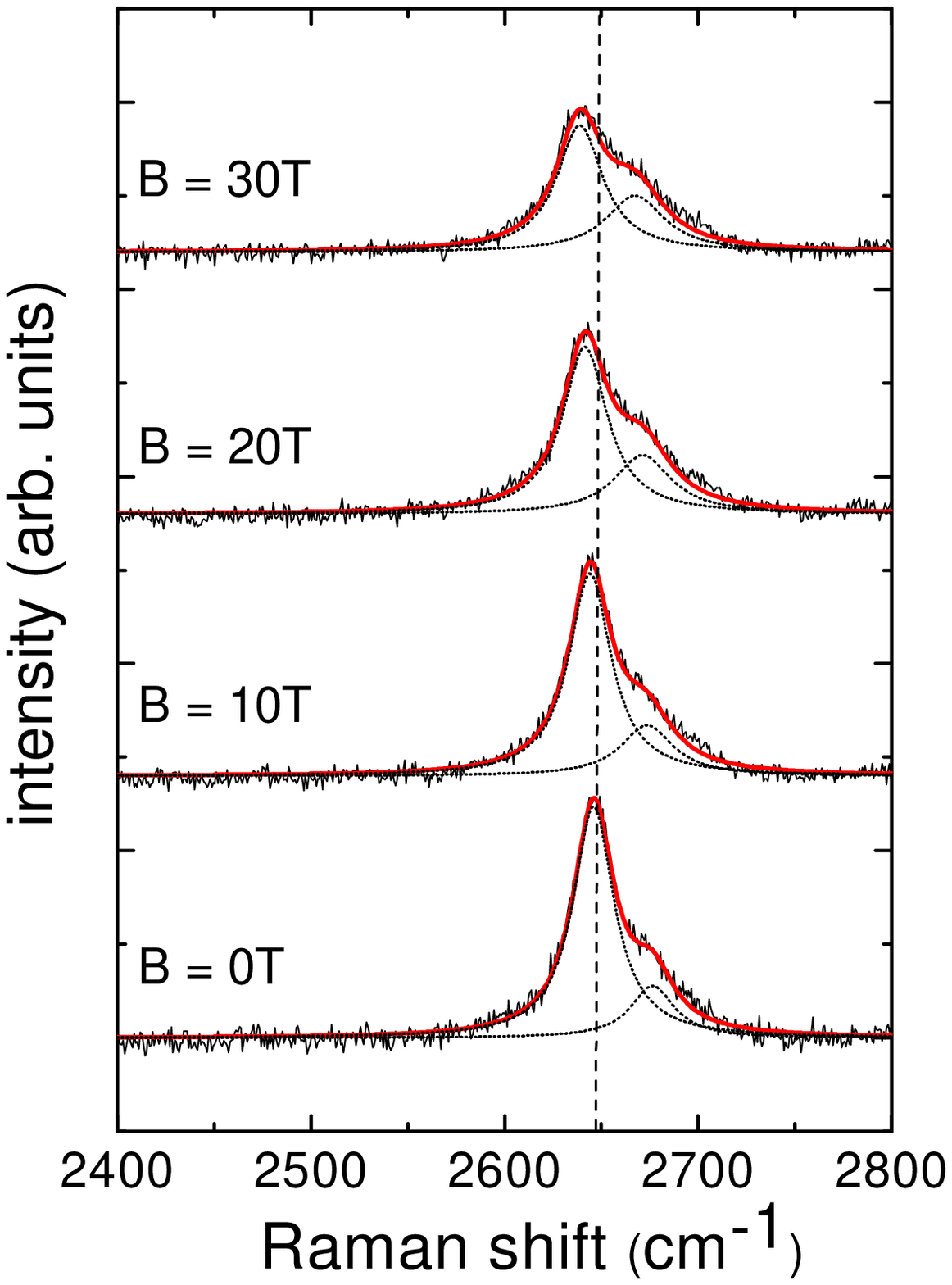}
\caption{(Color online) Experimental Raman scattering spectra measured at $T= 4.2\:\mbox{K}$ in the $2D$ band
range of energy at different values of the magnetic field (black
solid line), two Lorentzian fits (red solid line) and independent
components of the Lorentzian fits (black dotted lines). The
vertical black dashed line is a guide for the eyes.} \label{fig:spec}
\end{figure}

\begin{figure}
\includegraphics[width=1\linewidth,angle=0,clip]{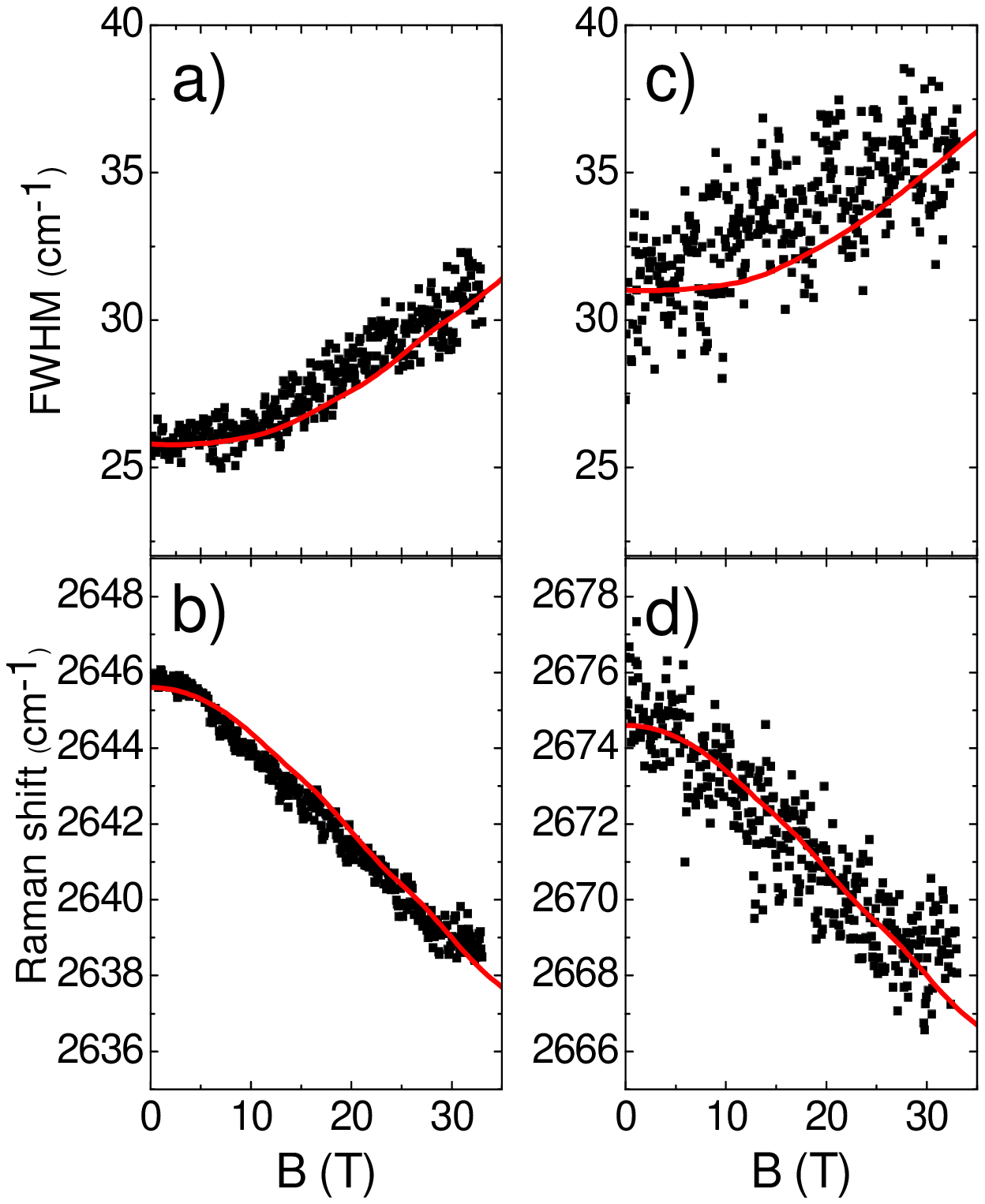}
\caption{(Color online) a) FWHM and b) Raman
shift of the low-energy component of the $2D$ band as a function of
the magnetic field. c) FWHM and d) Raman shift of the high-energy
component of the $2D$ band as a function of the magnetic field.
Solid red lines are calculated with our model (see text).}
\label{fig:param}
\end{figure}

Typical Raman scattering spectra of the $2D$ band measured at
$B=0, 10, 20$ and 30~T are presented in Fig.~\ref{fig:spec}. At zero
field, the $2D$ band is observed as a slightly asymmetric feature
composed of two Lorentzian contributions. The main contribution to
this feature is centered at 2646~cm$^{-1}$ with a full width at
half maximum (FWHM) of 26~cm$^{-1}$ and the second, weaker,
contribution is centered at 2675~cm$^{-1}$ with a FWHM of 31~cm$^{-1}$. As previously observed on similar
samples~\cite{Faugeras2008}, the energy at which the $2D$ band is
observed is increased, for a given excitation wavelength, with
respect to the energy at which the $2D$ band is expected for
exfoliated graphene flakes. The origin of this effect, probably
due to the peculiar type of stacking of the graphitic planes and
the resulting electronic interaction between these planes, is
still a matter of debate~\cite{Ni2008,Poncharal2008}. The origin
of the double component $2D$ band feature observed in our
experiment, different than the one observed in bulk graphite, is
still not clear. It could be due to Bernal-stacked inclusions
under the large laser spot. We note that similar line shapes have
also been observed in micro-Raman scattering measurements on
epitaxial graphene samples on SiC and were interpreted as
revealing graphene domains with different amount of
strain~\cite{Robinson2009}. When a magnetic field is applied
perpendicular to the graphene crystal plane, both of these two
components show a red shift and a broadening. We present in
Fig.~\ref{fig:param} the evolution of the Raman shift and of the FWHM
of both components of the $2D$ band with magnetic field. From
$B=0$ to 33~T, a shift of $\sim$ 8 cm$^{-1}$ and an increase of
20\% of the FWHM is observed.

\section{Discussion}

From the very beginning, we would like to stress that the observed
effect is quite unlikely to be related to the modification of the
spectrum of the emitted phonons (e.~g., by electron-phonon
interaction). Indeed, electrochemically top gated graphene
structures allowing to tune the Fermi level up to 800~meV (carrier
density up to $\sim4\times$10$^{13}$ cm$^{-2}$) have been produced
recently.\cite{Das2008,Das2009} Raman scattering experiments in
these highly doped structures showed that the evolution of the
$2D$ band central frequency with doping is moderate and
monotonous, and can be described by a change of the equilibrium
lattice parameter. Modification of the phonon dispersion due to
the coupling of phonons with low energy electronic excitations
across the Dirac point is noticeable near the $K$~point of the
phonon band structure (Kohn anomaly), while the phonons involved
in the $2D$~band are too far from the $K$~point to be affected by
the Kohn anomaly. This argument remains valid even in magnetic
fields used in our experiment. Indeed, the magnetic field mixes
wave vectors on the scale of the inverse magnetic length,
$1/l_B=\sqrt{eB/(\hbar{c})}\approx{0}.022\:\mbox{\AA}^{-1}$ at
$B=33\:\mbox{T}$. The wave vector of the phonons responsible for
the $2D$~peak, measured from the $K$~point, is
$q\approx{0}.24\:\mbox{\AA}^{-1}$ at 1.7 eV excitation. Thus,
$q\gg{1}/l_B$ and the Kohn anomaly remains inaccessible for these
phonons even at our highest fields. This is in striking contrast
with the main first-order Raman feature, the so-called $G$~band
due to the doubly degenerate $E_{2g}$ optical phonons at the
$\Gamma$ point of the phonon band structure. These phonons are
directly in the region of the Kohn anomaly, and their frequency
and lifetime are noticeably affected even by a moderate gate
voltage.\cite{Pisana2007,Yan2007} In strong magnetic fields, the
$G$~band exhibits a pronounced magneto-phonon
effect,\cite{Faugeras2009} as the phonon state is modified by the
coupling to electronic inter-Landau-level
transitions.\cite{Ando2007,Goerbig2007}

\begin{figure}
\includegraphics[width=1\linewidth]{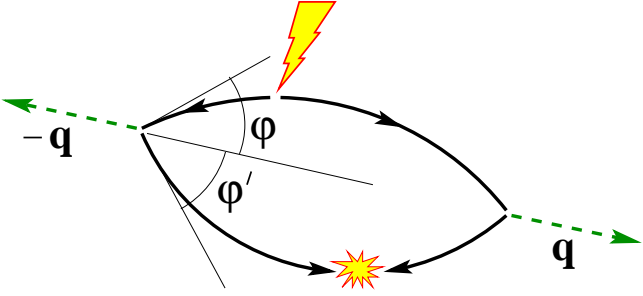}
\caption{(Color online) Schematic of the electron and hole motion during the
Raman scattering process. The lightening represents the incident
photon which creates the electron-hole pair. The solid arcs denote
the propagation of the electron and the hole in the magnetic
field. The flash represents the radiative recombination of the
electron-hole pair. The dashed arrows denote the emitted phonons.}
\label{fig:trajectories}
\end{figure}

In the following, we show that our observations are due to the
finite curvature of the trajectories of the photo-excited
electrons and holes in the magnetic field. The main idea is
illustrated in Fig.~\ref{fig:trajectories}. In the semi-classical
real-space picture of the Raman process,\cite{Basko2008} the
incident photon creates an electron and the hole with opposite
momenta at an arbitrary location within the laser spot. They
subsequently propagate along the classical trajectories, and emit
phonons. If they meet at some other location, again with opposite
momenta, they can recombine radiatively producing a scattered
photon. In the absence of the magnetic field, the trajectories are
straight lines, so that in order to meet at the same point with
opposite momenta and contribute to the Raman scattering signal,
the electron and the hole must necessarily be scattered
backwards during the phonon emission.\cite{Basko2008} This fixes the phonon
momentum~$\hbar{q}$ (measured from the $K$ or $K'$ point) as
$q=p+p'$, where $\hbar{p}=\hbar\win/(2v)$ and
$\hbar{p}'=\hbar(\win-2\omph)/(2v)$ are the electronic momenta
(also measured from the Dirac points) before and after the phonon
emission, $\win$~is the excitation frequency, $\omph$~is the
phonon frequency, and $v$~is the electronic velocity (the slope of
the Dirac cones). A slight spread in the phonon momentum due to
the quantum uncertainty, $|q-p-p'|\sim{2}\gamma/v$
($2\gamma$~being the electron inelastic scattering rate), gives a
contribution to the width of the $2D$~peak,
$\sim{2\gamma}(v_\mathrm{ph}/v)$, where $v_\mathrm{ph}=d\omph/dq$
is the phonon group velocity.\cite{Basko2008}

In a magnetic field, the electron and hole trajectories are no
longer straight lines but, because of the Lorentz force that acts
on charged particles in a magnetic field, they correspond to
circular cyclotron orbits. As a result, one can see from
Fig.~\ref{fig:trajectories} that (i)~phonons with smaller momenta,
$q=p\cos\varphi+p'\cos\varphi'$, can be emitted, and (ii)~since
each phonon can be emitted at an arbitrary instant in time, the
length of the arc describing the electron trajectory is random
[not exceeding the electron mean free path $v/(2\gamma)$], and so
are the angles~$\varphi,\varphi'$. Note that the frequencies of
the two emitted phonons remain equal even with an applied magnetic
field due to the symmetry under $C_2$ rotation around the axis
perpendicular to the crystal plane.~\footnote[1]{It is worth
noticing that only the relative arrangement of the two electron
trajectories and not their particular orientation with respect to
the crystal axis are relevant in our consideration. This
emphasizes the generality of our approach which is valid even if a
specific direction for electronic momentum could be assumed to
contribute to the 2D band Raman scattering signal in sp$^2$ carbon
materials\cite{Mafra2009}.} Tuning the magnetic field is, in this
sense, equivalent to changing the resonant conditions of the Raman
scattering process at a fixed excitation wavelength and allows
exploring part of the phonon band structure closer to the $K$
point. Fact~(i) results in the overall red shift of the Raman
peak, while fact (ii)~introduces an additional spread in~$q$, and
contributes significantly to the broadening of the peak as
observed through Raman scattering measurements.

To produce the observed effects, the electron and the hole do not
have to complete a full orbit. The probability of this latter event
vanishes as $e^{-2\pi{R}(2\gamma/v)}$, where
$R=p(eB/\hbar{c})^{-1}$ is the cyclotron radius. This implies that
the separation between the electronic Landau levels,
$\hbar\omega_c=\hbar{v}/R$, may still be smaller than their
broadening $2\hbar\gamma$ and, in our experiment, this is still
the case even at $B=30\:\mbox{T}$. In other words, even magnetic
fields far from the quantization limit can give a noticeable
effect. A closely related effect has been discussed in the context
of density-density response of a degenerate electron gas in a
non-quantizing magnetic field~\cite{Sedrakyan2007}.

To obtain some more quantitative information about the behavior of
the $2D$ peak in magnetic fields, we calculate the Raman matrix
element $\mathcal{M}(q)$ for a given phonon wave
vector~$\textit{q}$ in a clean graphene monolayer. It is
convenient to perform the calculation in the coordinate
representation, analogously to the calculation for the $D$~peak in
the vicinity of an edge\cite{Basko2009} (details of this calculation are presented in
the Appendix). An essential ingredient of the calculation is the semiclassical electronic
Green's function for graphene in a magnetic field which was
calculated in Ref.~\onlinecite{Ullmo2008}. Under the assumption
$\omega_c\ll(\gamma,\omph)\ll\win$, the result of the calculation
can be represented as:
\begin{eqnarray}
&&\mathcal{M}(q)\propto\int\limits_0^\infty{d}z\,\sqrt{z}\,
e^{-[i(q-2p)+2\gamma/v]z-i[p/(12R^2)]z^3}=\nonumber\\
&&\qquad=2(i\pi)^{3/2}\frac{R}{\sqrt{p}}\,
\frac{d}{du}\left[\Ai^2(u)-i\Ai(u)\Bi(u)\right],\qquad\label{matel=}\\
&&u\to\left(q-2p-\frac{2i\gamma}v\right)\left(\frac{R^2}{p}\right)^{1/3}.
\end{eqnarray}
For vanishing magnetic fields, the cyclotron radius $R\to\infty$
and the relation $\mathcal{M}(q)\propto(q-2p-2i\gamma/v)^{3/2}$,
which is Eq.~(64) of Ref.~\onlinecite{Basko2008}, is recovered.
For finite values of the magnetic field, the matrix element is
expressed in terms of Airy functions $\Ai(u)$ and
$\Bi(u)$~\cite{AiryBook}. Note also the
similarity with the expression for the polarization operator in
Ref.~\onlinecite{Sedrakyan2007}.

Since the density of the final phonon states is practically
$q$-independent\cite{PhononDOS},
$|\mathcal{M}(\win/v+\Omega/(2v_\mathrm{ph}))|^2$ describes the
Raman scattering intensity as a function of~$\Omega$, the Raman
shift measured with respect to the center of the $2D$ peak at zero
field. 
To illustrate the change of the peak shape in magnetic fields,
we present in Fig.~\ref{fig:noConv}(a)
$|\mathcal{M}(\win/v+\Omega/2v_\mathrm{ph})|^2$ for
$B=0,10,30\:\mbox{T}$, considering $\hbar\win=1.7\:\mbox{eV}$,
$\hbar{v}=7\:\mbox{eV}\cdot\mbox{\AA}$
($v=1.06\times{10}^8\:\mbox{cm/s}$),
$v_\mathrm{ph}/v=50\:\mbox{cm}^{-1}/\mbox{eV}$,
$\hbar\gamma=27\:\mbox{meV}$. This value for $\gamma$ is chosen
in order to reproduce the observed shift of the peak maximum for
increasing magnetic fields. However, the resulting FWHM for the
$2D$ band is about twice smaller than the one observed
experimentally at zero field. Most likely, this indicates
the presence of an additional broadening mechanism besides the
electronic scattering. As we cannot determine the nature of this
mechanism, we model it by introducing phenomenologically an
additional broadening of the $2D$ band through a convolution of
$|\mathcal{M}|^2$ with a Gaussian curve of width $\sigma$:
\begin{equation}
I_{2D}(\Omega)\propto\int\limits_{-\infty}^{\infty}
\frac{d\Omega'}{\sqrt{2\pi}\sigma}\,e^{-\frac{(\Omega'-\Omega)^2}{2\sigma^2}}
\left|\mathcal{M}\!\left(\frac\win{v}+\frac{\Omega'}{2v_\mathrm{ph}}\right)\right|^2.
\label{gaussbroad=}
\end{equation}
There is no simple relation between the FWHM of the original
$|\mathcal{M}|^2$ peak ($8\gamma(v_{ph}/v)\sqrt{2^{2/3}-1}$ at
zero field), the FWHM of the broadening Gaussian
($2\sigma\sqrt{\ln{4}}$), and the FWHM of the resulting peak. The
result of the convolution for $\sigma=9.5\:\mbox{cm}^{-1}$ is
shown in Fig.~\ref{fig:noConv}(b).

\begin{figure*}
\includegraphics[width=1\linewidth]{TheoryPeak.eps}
\caption{(a)~$|\mathcal{M}|^2$ as given by Eq.~(\ref{matel=}) for
$\gamma=27\:\mbox{meV}$. (b)~convolution of $|\mathcal{M}|^2$ with
a $9.5\:\mbox{cm}^{-1}$-wide Gaussian curve as given by
Eq.~(\ref{gaussbroad=}). Solid, dashed, and dotted lines
correspond to $B=0,10,30\:\mbox{T}$, respectively. The Raman shift
is measured with respect to the center of the peak at zero field.}
\label{fig:noConv}
\end{figure*}

\begin{figure}
\includegraphics[width=0.8\linewidth,angle=0,clip]{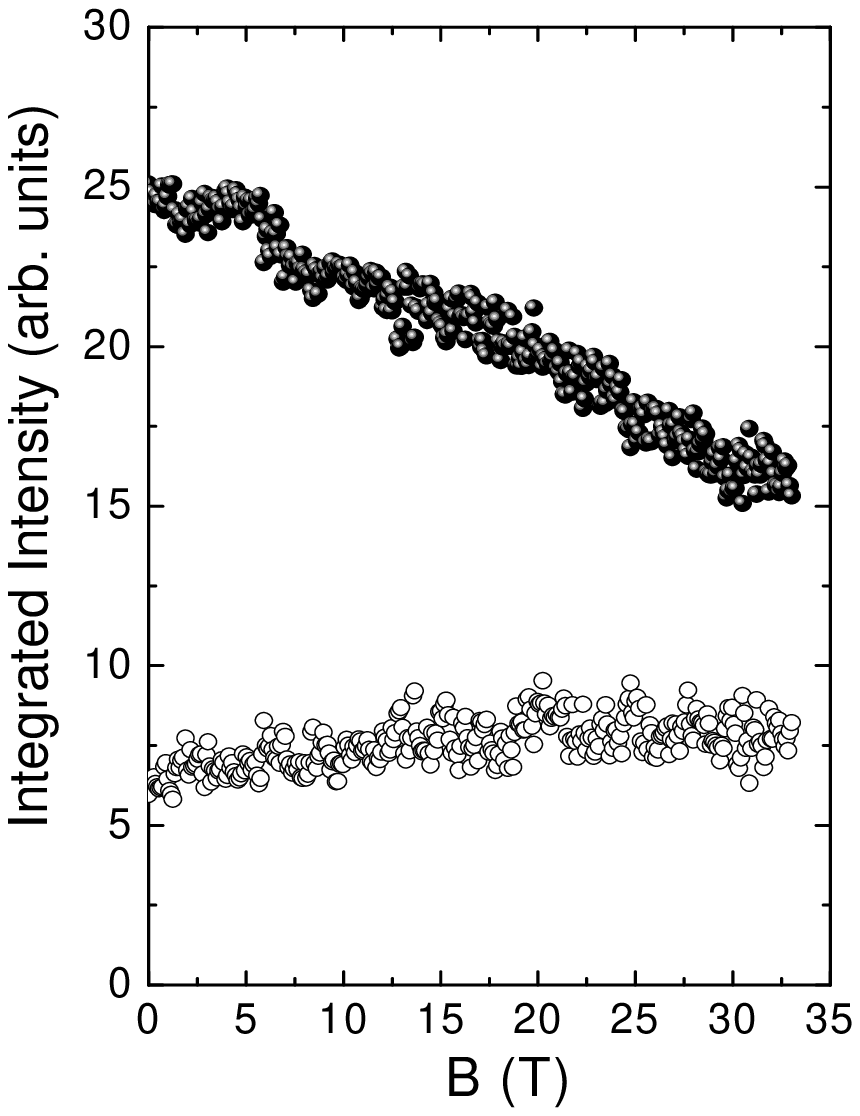}
\caption{Integrated intensity of the 2D band low energy component (black
dots) and of the 2D band high energy component (open circles) as a
function of the magnetic field.} \label{fig:intensity}
\end{figure}

In the following, we assume $\gamma$ and $\sigma$ to be
independent of the magnetic field. This approximation is
reasonable as long as the magnetic field is
non-quantizing.\cite{noBdep} At $B=30\:\mbox{T}$, electrons with
energy $\epsilon=0.85\:\mbox{eV}$  measured from the Dirac point,
have a cyclotron frequency
$\omega_c=(v^2\hbar/\epsilon)(eB/\hbar{c})$ of about
$27\:\mbox{meV}$, so the strongest magnetic fields in this
experiment seem to be close to the limits of validity of this
approximation. Beyond this regime, Landau quantization of the
electronic spectrum should manifest itself as oscillations of the
peak intensity due to periodic modifications of the resonance
conditions with increasing magnetic field, while in the experiment
no such oscillations are observed.

The results of this calculation in terms of Raman shift and of
FWHM of the peak, as described by Eq.~(\ref{gaussbroad=}), are
presented as red solid lines in Fig.~\ref{fig:param} (a)--(d)
for the two components of the observed 2D band feature. The
following parameters were adjusted: (i)~the central frequencies
of the two components at zero field, giving the overall vertical
offset for the curves in Fig.~\ref{fig:param}~(b),~(d);
(ii)~$\sigma=9.5\:\mbox{cm}^{-1}$ and $\sigma=11.9\:\mbox{cm}^{-1}$
were taken for the low-frequency and high-frequency components,
respectively, in order to reproduce the FWHM at zero field, in
combination with (iii)~$\hbar\gamma=27\:\mbox{meV}$ which determines
\emph{all four} slopes in Fig.~\ref{fig:param} (a)--(d).
Clearly, without precise knowledge of the origin of the
two $2D$~band components we cannot account for the slight difference
in their zero-field widths. Nevertheless, the fact that the
calculation reproduces the red shift of the $2D$ band energy as a
function of the magnetic field which is quadratic for fields up to
5--10~T and linear at higher fields, with a single value of~$\gamma$
for both components, shows that the electrons have the same dynamics
in the parts of the sample, responsible for the two components.

Besides, the deduced value of $\hbar\gamma=27\:\mbox{meV}$ is in
good agreement with that deduced from the doping dependence of the
$2D$~peak intensity in exfoliated graphene\cite{BPF2009,Casiraghi2009}
(given the energy dependence of~$\gamma$ and the fact that the latter measurements were
performed at higher excitation energy, the electron scattering in
those samples  is somewhat weaker than in ours). We also note that
the found value $\hbar\gamma=27\:\mbox{meV}$ is in reasonable
agreement with the line width of electronic transitions in high
magnetic fields measured in this range of energy on similar
samples.\cite{Plochocka2008} This fact is remarkable because those
measurements were performed in a quantizing magnetic field, and
\emph{a priori} the scattering rate does not have to be the
same.\cite{noBdep}

From Eq.~(\ref{matel=}) it can be seen that the integral
$\int|\mathcal{M}(q)|^2dq$ does not depend on the magnetic field.
(It is sufficient to integrate over~$q$ first, and then over~$z$;
the magnetic field enters only through~$R$, which drops out.) This
means that under the assumption of a constant phonon density of
states, the frequency-integrated intensity (the area under the
peak) of the $2D$~band should not depend on the magnetic field.
The experimental intensities for the two components are plotted on
Fig.~\ref{fig:intensity}. Only the integrated intensity of the
high-frequency component is field independent, while an overall
decrease of $\sim{35}\%$ of the integrated intensity of the
low-frequency component is observed over the range of~$B$ between
0 and 33~T. Does it mean that the phonon density of states
decreases stronger for the lower-frequency component? Again,
without precise knowledge of the exact nature of the two
components it is hard to give an explanation for their different
behavior.

\section{conclusions}

In conclusion, we have studied, both experimentally and
theoretically, the evolution of the $2D$ band of MEG structures in
intense magnetic fields. We observe a red shift and a broadening
of the $2D$ band as the magnetic field is increased. We have
modelled this effect using a semi-classical picture in which the
Lorentz force induced by the magnetic field on the photo-created
electron-hole pairs, curves the carriers trajectories. This leads
to a decrease of the emitted optical phonon momentum and a
broadening of the $2D$ band as observed through Raman scattering
spectroscopy. This model enables us to extract the value of the
electronic scattering rate. From this we conclude that about half
of the $2D$~band width at zero field is due to electronic
scattering while the origin of the other half remains to be
clarified. Finally, we note that the observed effect of the
magnetic field on the $2D$~peak is not specific to monolayer
graphene; it should be analogous for multilayer graphene and
graphite.

\begin{acknowledgments}
Part of this work has been supported by EC MTKD-CT-2005-029671,
EuromagNetII, PICS-4340, ANR-08-JCJC-0034-01 and ANR-06-NANO-019
projects. P.K. is financially supported by the EU under FP7,
contract no. 221515 `MOCNA'. M. A. thanks the Cariplo Foundation
(project QUANTDEV), MICINN (Project MOSAICO) and JCYL SA052A07 for
support.
\end{acknowledgments}

\appendix

\section{Quasiclassical calculation of the two-phonon Raman matrix
element in a uniform magnetic field}

\subsection{Classical action for a Dirac particle}

The two-phonon Raman matrix element is given by a loop of four
electronic Green's functions. The latter will be taken in the
quasiclassical approximation. The most important ingredient of the
quasiclassical Green's function is the classical action. Thus, we
first discuss the classical action, then we give the explicit
expression for the electronic Green's function, and finally, we
perform the calculation of the Raman matrix element. In fact, the
first two steps have been already made by Carmier and
Ullmo \cite{Ullmo2008}, but we include them here for the sake of
completeness of the presentation, and to fix the notations.

We start with the classical equation of motion of a charge in the
magnetic field~$\vec{B}$:
\begin{equation}
\frac{d\vec{p}}{dt}=\frac{e}c\,\vec{v}\times\vec{B}
\label{Lorentz=}
\end{equation}
For a Dirac electron with energy~$\ep$ the velocity
$\vec{v}(t)=v\vec{n}(t)$ and momentum
$\vec{p}(t)=(\ep/v)\vec{n}(t)$ are expressed in terms of a unit vector,
$|\vec{n}|=1$. [If instead of electrons with positive and negative
energies one prefers to work with positive-energy electrons and holes,
then $\vec{p}(t)=|\ep/v|\vec{n}(t)$, so the action given below should
be multiplied by $\sign\ep$].
Eq.~(\ref{Lorentz=}) can be obtained by variation of the action
functional
\begin{equation}
S[\vec{r}(t)]=\int{d}t\left(\frac{\ep}v\,|\dot{\vec{r}}|
+\frac{e}{c}\,\vec{A}(\vec{r})\cdot{\dot{\vec{r}}}\right)
\end{equation}
keeping the ends $\vec{r}_1,\vec{r}_2$ of the trajectory fixed.
Indeed, upon integration by parts
\begin{eqnarray}
\delta{S}&=&\int{d}t\left[-\frac\ep{v}\,\frac{d}{dt}\frac{\dot{x}_i}{|\dot{\vec{r}}|}
+\frac{e}c\,\dot{x}_j
\left(\frac{\partial{A}_j}{\partial{x}_i}-\frac{\partial{A}_i}{\partial{x}_j}\right)\right]\delta{x}_i
\nonumber\\
&&{}+\vec{p}_2\cdot\delta\vec{r}_2
-\vec{p}_1\cdot\delta\vec{r}_1,
\end{eqnarray}
where indices $i,j=x,y,z$ label the Cartesian components, and the momentum
is defined as
\begin{equation}
\vec{p}\equiv\frac{\ep}v\,\frac{\dot{\vec{r}}}{|\dot{\vec{r}}|}
+\frac{e}{c}\,\vec{A}(\vec{r}).
\end{equation}
If we now define the function $S(\vec{r},\vec{r}')$ as the action
on the classical trajectory, corresponding to the motion from~$\vec{r}'$
to~$\vec{r}$ according to Eq.~(\ref{Lorentz=}), we obtain
$\vec\nabla{S}=\vec{p}$ at the end of the trajectory.
Thus, this function satisfies the Hamilton-Jacobi equation
\begin{equation}
\left|\vec\nabla{S}(\vec{r},\vec{r}')-\frac{e}c\,\vec{A}(\vec{r})\right|=
\left|-\vec\nabla'{S}(\vec{r},\vec{r}')-\frac{e}c\,\vec{A}(\vec{r}')\right|=
\frac{|\ep|}v.
\end{equation}
Also, $\partial{S}/\partial\ep$ is equal to the time it takes to
go along the trajectory.

\begin{figure}
\includegraphics[width=0.8\linewidth,angle=0,clip]{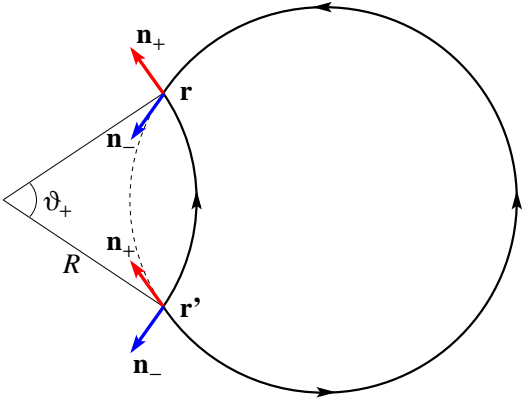}
\caption{\label{fig:Larmor} (Color online) Two classical trajectories connecting
the points $\vec{r}$, $\vec{r}'$.}
\end{figure}

In a uniform magnetic field, $\vec{n}$ is
precessing with a constant frequency $\omega=-(eB/c)(v^2/\ep)$,
and the trajectories are circles of the radius $R=v/|\omega|$.
Two particular points $\vec{r}$ and $\vec{r}'$ can be connected either
by two trajectories corresponding to short and long arcs (plus an
integer number of full rotations), or by no trajectories at all if
the distance between the points is greater than the circle diameter,
$|\vec{r}-\vec{r}'|>2R$.
Let us assume $\vec{B}$ to be along the $z$ axis and choose the gauge
$\vec{A}(\vec{r})=(B/2)[\vec{e}_z\times\vec{r}]$.
The action along the short/long arc is given by
\begin{eqnarray}\label{Sclpm=}
&&S_\pm(\vec{r},\vec{r}')=\frac{\ep{R}}{2v}
\left[\vartheta_\pm(\vec{r},\vec{r}')
+\sin\vartheta_\pm(\vec{r},\vec{r}')\right]
\nonumber\\ &&\qquad\qquad{}
-\frac{eB}{2c}\,[\vec{r}\times\vec{r}']_z,\\
&&\vartheta_+(\vec{r},\vec{r}')=2\arcsin\frac{|\vec{r}-\vec{r}'|}{2R},\\
&&\vartheta_-(\vec{r},\vec{r}')=2\pi-2\arcsin\frac{|\vec{r}-\vec{r}'|}{2R}.
\end{eqnarray}
Here $\vartheta_\pm(\vec{r},\vec{r}')$ is the angular size of the
short/long arc. We also introduce
$\vartheta_\pm^{(j)}(\vec{r},\vec{r}')=2\pi{j}+\vartheta_\pm(\vec{r},\vec{r}')$.
We denote by $\vec{n}_\pm$ the unit tangent vector to the short/long
arc at the point $\vec{r}$ (Fig.~\ref{fig:Larmor}):
\begin{eqnarray}
\vec{n}_\pm(\vec{r},\vec{r}')&=&\left[\vec{e}_z\times\frac{\vec{r}-\vec{r}'}{|\vec{r}-\vec{r}'|}\right]
\sin\frac{\vartheta_\pm(\vec{r},\vec{r}')}2\sign\omega\nonumber\\
&&{}+\frac{\vec{r}-\vec{r}'}{|\vec{r}-\vec{r}'|}
\cos\frac{\vartheta_\pm(\vec{r},\vec{r}')}2.
\end{eqnarray}
The tangent determines the direction of the kinematic momentum:
\begin{eqnarray}
&&\vec\nabla{S}_\pm(\vec{r},\vec{r}')-\frac{e}c\vec{A}(\vec{r})
=\frac\ep{v}\,\vec{n}_\pm(\vec{r},\vec{r}'),\\
&&-\vec\nabla'{S}_\pm(\vec{r},\vec{r}')-\frac{e}c\vec{A}(\vec{r}')
=-\frac\ep{v}\,\vec{n}_\mp(\vec{r},\vec{r}').
\end{eqnarray}

\subsection{Quasiclassical Green's function}

Now we discuss the Green's function for the Dirac equation,
\begin{eqnarray}
\left[\ep+i\hbar\gamma_\ep\sign\ep
+v\vec\Sigma\cdot\left(i\hbar\vec\nabla+\frac{e}c\,\vec{A}\right)\right]
G(\vec{r},\vec{r}')\nonumber\\
=\unitmatrix\,\delta(\vec{r}-\vec{r}').
\label{Dirac=}
\end{eqnarray}
Here $\vec\Sigma=(\Sigma_x,\Sigma_y)$ are the pseudospin matrices
in the sublattice space, $\unitmatrix$~is the unit matrix,
and $\gamma_\ep>0$ is the half of the electron or hole inelastic
scattering rate, introduced phenomenologically (see Sec. IVC of
Ref.~\onlinecite{Basko2008} for the discussion of its role in
Raman scattering). The factor $\sign\ep$ corresponds to the
chronologically ordered Green's function.

The quasiclassical Green's function is represented as a sum over
all trajectories of the classical motion from~$\vec{r}'$ to~$\vec{r}$,
i.~e., sum over the short and long arcs, ``$\pm$'', and over the
number of rotations $j=0,1,\ldots$:
\begin{eqnarray}
&&G(\vec{r},\vec{r}')=
\sum_{j=0}^\infty\left[G_+^{(j)}(\vec{r},\vec{r}')+
G_-^{(j)}(\vec{r},\vec{r}')\right],\label{Gfull=}\\
&&G_\pm^{(j)}(\vec{r},\vec{r}')=e^{\pm{i}\pi/4}\sign\ep
\sqrt{\frac{|\ep|/(\hbar{v})^3}
{2\pi{R}|\sin\vartheta_\pm^{(j)}(\vec{r},\vec{r}')|}}\times\nonumber\\
&&\qquad\qquad{}\times
\psi_{\vec{n}_\pm(\vec{r},\vec{r}')}
\psi^\dagger_{-\vec{n}_\mp(\vec{r},\vec{r}')}e^{-i(eB/2\hbar{c})[\vec{r}\times\vec{r}']_z}
\times\nonumber\\
&&\qquad\qquad{}\times
e^{i[\vartheta_\pm^{(j)}(\vec{r},\vec{r}')+\sin\vartheta_\pm^{(j)}(\vec{r},\vec{r}')]
|\ep|R/(2\hbar{v})}\times\nonumber\\
&&\qquad\qquad{}\times
e^{-\vartheta_\pm^{(j)}(\vec{r},\vec{r}')\gamma_\ep{R}/v}.\label{Gjpm=}
\end{eqnarray}
The imaginary part of the argument of the exponential is just the
classical action discussed in the previous subsection.
Replacement $\ep\to|\ep|$ is required by the analytical properties
of the chronologically ordered Green's function (otherwise it would
correspond to the retarded Green's function).
The damping factor $e^{-\gamma_\ep\vartheta{R}/v}$ is determined by
the total length of the corresponding arc $\vartheta{R}$, including
$2\pi{j}R$ from $j$~full rotations. The eigenvector $\psi_\vec{n}$ of
$(\vec{n}\cdot\vec\Sigma)\psi_\vec{n}=\psi_\vec{n}\sign\ep$,
$|\vec{n}|=1$, can be written in terms of the polar angle
$\varphi_\vec{n}$ of the unit vector~$\vec{n}$ (we use the basis
where $\Sigma_x,\Sigma_y$ are represented by Pauli matrices),
\begin{equation}
\psi_\vec{n}=\frac{1}{\sqrt{2}}
\left(\begin{array}{c}e^{-i\varphi_\vec{n}/2}\sign\ep
\\ e^{i\varphi_\vec{n}/2}\end{array}\right).
\end{equation}
Thus defined, $\psi_\vec{n}$ is not a single-valued function: when
$\vec{n}$~is rotated by $2\pi$, $\psi_\vec{n}$~acquires a minus sign.
We can fix the sign by requiring that
at $\vec{r}\to\vec{r}'$,
$\psi_{\vec{n}_+}=-\psi_{-\vec{n}_-}=\psi_{\vec{r}-\vec{r}'}$,
$\psi_{\vec{n}_-}=\psi_{-\vec{n}_+}=\psi_{\vec{r}'-\vec{r}}$,
and they evolve continuously during the motion along the circle.

We make several further remarks concerning the above expression for
$G(\vec{r},\vec{r}')$, Eqs.~(\ref{Gfull=}),~(\ref{Gjpm=}).

(i)~For $\vec{r}\to\vec{r}'$ $G(\vec{r},\vec{r}';\ep)$ reduces
to the expression for the quasiclassical Green's function for $B=0$,
\begin{eqnarray}
G_0(\vec{r}-\vec{r}';\ep)&=&
-\frac{e^{i\pi/4}}2\left(\unitmatrix\sign\ep
+\frac{\vec{r}-\vec{r}'}{|\vec{r}-\vec{r}'|}\cdot\vec\Sigma\right)
\nonumber\\
&&\times\sqrt{\frac{|\ep|/(\hbar{v})^3}{2\pi|\vec{r}-\vec{r}'|}}\,
e^{(i|\ep|/\hbar-\gamma_\ep)|\vec{r}-\vec{r}'|/v},\nonumber\\
\end{eqnarray}
valid at $|\ep||\vec{r}-\vec{r}'|/(\hbar{v})\gg{1}$.

(ii)~The difference between the exact Green's function and
$G(\vec{r},\vec{r}';\ep)$ given by Eqs.~(\ref{Gfull=}),~(\ref{Gjpm=})
is of higher order in~$\hbar$. This can be checked explicitly
by calculating the gradients:
\begin{eqnarray}
&&d\varphi_{\vec{n}_\pm(\vec{r},\vec{r}')}=
-\frac{[\vec{n}_\mp\times{d}\vec{r}]_z}{R\sin\vartheta_\pm}
=\frac{[\vec{n}_\mp\times{d}\vec{r}']_z}{R\sin\vartheta_\pm},\\
&&\vec\nabla\sin\vartheta_\pm=\frac{1}R\,
\frac{\cos\vartheta_\pm}{\cos(\vartheta_\pm/2)}\frac{\vec{r}-\vec{r}'}{|\vec{r}-\vec{r}'|},\\
&&\vec\nabla\psi_{\vec{n}_+}=
\frac{1}{2i}\,\Sigma_z\psi_{\vec{n}_+}\vec\nabla\varphi_{\vec{n}_+}\nonumber\\
&&=\frac{1}{2i}\,\Sigma_z\psi_{\vec{n}_+}
\frac{[\vec{e}_z\times\vec{n}_-]}{R\sin\vartheta_+},\\
&&(\vec\Sigma\cdot\vec\nabla)\psi_{\vec{n}_+}=
-\frac{(\vec{n}_-\cdot\Sigma)\psi_{\vec{n}_-}}{2R\sin\vartheta_+},
\end{eqnarray}
so that at $\vec{r}\neq\vec{r}'$
\begin{eqnarray}
&&[\ep+i\gamma_\ep\sign\ep
+i\hbar{v}\vec\Sigma\cdot\vec\nabla]G_+^{(j)}(\vec{r},\vec{r}';\ep)=\nonumber\\
&&=-\frac{i\hbar{v}}{R\cos^2(\vartheta_+/2)}\,[\vec{n}_+\times\vec\Sigma]_z
G_+^{(j)}(\vec{r},\vec{r}';\ep).
\end{eqnarray}
It is proportional to the eigenvector corresponding to the energy $-\ep$;
thus, it can be compensated by a correction to the pre-exponential
factor which does not vanish when acted upon by
$\ep+iv\vec\Sigma\cdot(\vec\nabla{S}-ie\vec{A}/c)$, and thus is of the
next order in~$\hbar$:
\begin{equation}
\delta{G}_+^{(j)}(\vec{r},\vec{r}';\ep)
=\frac{i\hbar{v}}{2\ep{R}\cos^2(\vartheta_+/2)}\,[\vec{n}_+\times\vec\Sigma]_z
G_+^{(j)}(\vec{r},\vec{r}';\ep).
\end{equation}
$\delta{G}_+^{(j)}$ is smaller than $G_+^{(j)}$ by the dimensionless
parameter $\hbar{v}/(\ep{R})$. However, it has a stronger divergence
at $|\vec{r}-\vec{r}'|\to{2}R$; this is the manifestation of the usual
breakdown of the quasiclassical approximation in the vicinity of the
classical turning point.

(iii)
To fix the relative phases of $G_+^{(j)}$ and $G_-^{(j)}$ we choose
a given circular trajectory and consider the evolution of
$G(\vec{r},\vec{r}';\ep)$ as $\vec{r}$ moves along the circle. Namely, as
$\vec{r}$ passes the turning point at $|\vec{r}-\vec{r}'|=2R$,
$G_+^{(j)}(\vec{r},\vec{r}';\ep)$ should transform into
$G_-^{(j)}(\vec{r},\vec{r}';\ep)$.
In the vicinity of the turning point, $|\vec{r}-\vec{r}'|=2R(1+z)$,
$z\ll{1}$, we have $\vartheta_\pm\approx\pi\mp\sqrt{-2z}$.
Thus, $\vartheta_\pm$ can be viewed as two branches of the
same analytical function, and $\vartheta_+\to\vartheta_-$ when
$z$~makes a circle around $z=0$ in the complex plane. Since
$\vartheta_\pm+\sin\vartheta_\pm\approx\pi\mp(-2z)^{3/2}/6$,
in order to get a decaying exponential in the classically forbidden
region $|\vec{r}-\vec{r}'|>2R$ (the positive semiaxis of~$z$),
the circle must be counterclockwise. Thus,
\begin{equation}
\frac{e^{i\pi/4}}{\sqrt{\sin\vartheta_+^{(j)}}}\to
\frac{e^{-i\pi/4}}{\sqrt{|\sin\vartheta_-^{(j)}|}},
\end{equation}
precisely as in Eq.~(\ref{Gjpm=}).

(iv)
To match the phases of $G_-^{(j)}$ and $G_+^{(j+1})$, we consider
$\vec{r}\to\vec{r}'$ and require that
$G_-^{(j)}(\vec{r},\vec{r}';\ep)+G_+^{(j+1)}(\vec{r},\vec{r}';\ep)$ satisfies
the homogeneous Dirac equation. Indeed, the $\delta$-function term is
produced by $G_+^{(0)}$ component, see remark~(i).
At $\vec{r}\to\vec{r}'$ the Green's function satisfying the homogeneous
Dirac equation can be can be constructed in the $B\to{0}$ limit.
Let us introduce $\bar{G}_0(\vec{r}-\vec{r}';\ep)$, the Green's
function with analytical properties opposite to those of~$G_0$:
\begin{eqnarray}
\bar{G}_0(\vec{r}-\vec{r}';\ep)&=&
\frac{e^{-i\pi/4}}2\left(-\unitmatrix\sign\ep
+\frac{\vec{r}-\vec{r}'}{|\vec{r}-\vec{r}'|}\cdot\vec\Sigma\right)
\nonumber\\
&&\times\sqrt{\frac{|\ep|/(\hbar{v})^3}{2\pi|\vec{r}-\vec{r}'|}}\,
e^{(-i|\ep|/\hbar+\gamma_\ep)|\vec{r}-\vec{r}'|/v}.\nonumber\\
\end{eqnarray}
It also satisfies Eq.~(\ref{Dirac=}), as seen from the antisymmetry
of the Dirac operator with respect to the simultaneous change
$\ep\to-\ep$, $i\to-i$, $\gamma_\ep\to-\gamma_\ep$. Thus,
$G_0(\vec{r}-\vec{r}';\ep)-\bar{G}_0(\vec{r}-\vec{r}';\ep)$ represents the
sought solution, corresponding to the flux of particles
focusing at the point $\vec{r}'$. Hence, the correspondence is
$G_-^{(j)}(\vec{r}\to\vec{r}';\ep)\sim-\bar{G}_0(\vec{r}-\vec{r}';\ep)$
(converging wave),
$G_+^{(j+1)}(\vec{r}\to\vec{r}';\ep)\sim{G}_0(\vec{r}-\vec{r}';\ep)$
(outgoing wave), which produces precisely the combination of signs
as in Eq.~(\ref{Gjpm=}).

Remarks (iii) and (iv) can be restated in very simple terms. Upon
every half-rotation $\sin\vartheta$ changes sign, so the square root
produces a factor of $e^{i\pi/2}$. In addition, upon a full rotation
$\psi_\vec{n}$ acquires the Berry phase $e^{i\pi}$. Hence, the sum over
$j$ in Eqs.~(\ref{Gfull=}), (\ref{Gjpm=}) reduces to a simple geometric
progression,
\begin{equation}
\sum_{j=0}^\infty e^{j\pi(i|\ep|/\hbar-2\gamma)R/v}=
\frac{1}{1-e^{\pi(i|\ep|/\hbar-2\gamma)R/v}},
\end{equation}
which determines the poles corresponding to $\pi\ep{R}/\hbar{v}$
being an integer multiple of $2\pi$, i.~e., Bohr-Sommerfeld quantization
rule. It gives the exact expression for the Landau levels,
$|\ep_n/v|=\sqrt{2n|e\hbar{B}/{c}|}$.

\subsection{Two-phonon Raman matrix element}

Here we consider the matrix element
$\mathcal{M}(\vec{q};\varphi_{\mathrm{in}},\varphi_{\mathrm{out}})$ for
the transition from the initial state, corresponding  to the incident
photon with frequency~$\win$ polarized at the angle $\varphi_\mathrm{in}$
to the $x$~axis, and no phonons, to the final state, corresponding to the
scatterd photon with frequency $\win-2\omega_\vec{q}$ polarized at the
angle $\varphi_\mathrm{out}$ to the $x$~axis, and two phonons with momenta
$\vec{q},-\vec{q}$ and frequencies $\omega_\vec{q}$ ($\vec{q}$ is measured
from the $K$ or $K'$ point, and for the phonons in the two valleys the
frequencies $\omega^{(K)}_\vec{q}=\omega^{(K')}_{-\vec{q}}$ due to the
time-reversal symmetry; we denote $\omega^{(K)}_\vec{q}\equiv\omega_\vec{q}$
and omit the valley index).
Then, for unpolarized excitation and detection, the frequency-resolved
intensity $I(\Omega)$ is given by
\begin{eqnarray}
I(\Omega)&\propto&\int\limits_0^{2\pi}
\frac{d\varphi_\mathrm{in}}{2\pi}
\frac{d\varphi_\mathrm{out}}{2\pi}
\int\frac{d^2\vec{q}}{(2\pi)^2}\nonumber\\&&{}\times
|\mathcal{M}(\vec{q};\varphi_{\mathrm{in}},\varphi_{\mathrm{out}})|^2
\delta(\Omega-2\omega_\vec{q}).
\end{eqnarray}
The matrix element is given by the loop of four Green's function with
electron-photon and electron-phonon vertices (here we write it in the
coordinate representation)~\cite{Basko2008}:
\begin{eqnarray}
\mathcal{M}(\vec{q};\varphi_{\mathrm{in}},\varphi_{\mathrm{out}})
&\propto&\int\frac{d\ep}{2\pi}\int
{d}^2\vec{r}_2\,{d}^2\vec{r}_1\,{d}^2\vec{r}_0\,{d}^2\vec{r}_1'
\nonumber\\ &&{}\times
\Tr\left\{(\Sigma_x\cos\varphi_\mathrm{out}+\Sigma_y\sin\varphi_\mathrm{out})
\right.\nonumber\\&&{}\times
G(\vec{r}_2,\vec{r}_1;\ep+\win/2-\omega_\vec{q})
\nonumber\\&&{}\times
\Sigma_ze^{-i\vec{q}\vec{r}_1}G(\vec{r}_1,\vec{r}_0;\ep+\win/2)
\nonumber\\&&{}\times
(\Sigma_x\cos\varphi_\mathrm{in}+\Sigma_y\sin\varphi_\mathrm{in})\,
\nonumber\\&&{}\times
G(\vec{r}_0,\vec{r}_1';\ep-\win/2)\,\Sigma_ze^{i\vec{q}\vec{r}_1'}
\nonumber\\&&{}\times\left.
G(\vec{r}_1',\vec{r}_2;\ep-\win/2+\omega_{\vec{q}})\right\}.\nonumber\\
\end{eqnarray}
We will assume $\gamma_{\win/2}{R}/v\gg{1}$ (non-quantizing field),
then the probability of a half- or full rotation is exponentially small,
and only the $G_+^{(0)}$ contribution to the Green's functions remains.
The integral over $\ep$ is dominated by small $|\ep|\sim\gamma_{\win/2}$,
so the energy arguments of the Green's functions can be assumed to have
definite signs. Thus, the imaginary part of the exponent can be written as
\begin{eqnarray}
&&S(\vec{r}_2,\vec{r}_1;\win/2-\omega_\vec{q}+\ep)-\vec{q}\vec{r}_1
+S(\vec{r}_1,\vec{r}_0;\win/2+\ep)\nonumber\\
&&{}+S(\vec{r}_0,\vec{r}_1';\win/2-\ep)+\vec{q}\vec{r}_1'\nonumber\\
&&{}
+S(\vec{r}_1',\vec{r}_2;\win/2-\omega_\vec{q}+\ep),
\label{exponent=}
\end{eqnarray}
where the action $S$ is that given by Eq.~(\ref{Sclpm=}), with the
``+'' subscript omitted and the energy argument explicitly introduced.

The spatial integration is performed in the stationary point approximation.
Namely, in the whole 8-dimensional space
$(\vec{r}_0,\vec{r}_1,\vec{r}_1',\vec{r}_2)$
we separate a manifold on which expression~(\ref{exponent=}) is stationary.
As $\vec\nabla{S}$ is just the classical momentum, this manifold corresponds
to joining the classical arcs connecting the pairs of points in order to
satisfy momentum conservation at each point
$\vec{r}_0,\vec{r}_1,\vec{r}_1',\vec{r}_2$.
Then the integration over the deviations from this manifold is performed
in the Gaussian approximation, while the integration over the manifold
itself has to be done more carefully. This procedure is fully analogous to
that used in Sec.~VIB of Ref.~\onlinecite{Basko2009} for the edge-assisted
Raman scattering.

The integration over~$\ep$ will be performed by expanding the actions to
the linear order, e.~g.
\begin{equation}\begin{split}
&S(\vec{r}_1,\vec{r}_0;\win/2+\ep)\to\\ &\to S(\vec{r}_1,\vec{r}_0;\win/2)
+\left.\frac{\partial{S}(\vec{r}_1,\vec{r}_0;\win/2+\ep)}{\partial\ep}\right|_{\ep=0}\ep,
\end{split}\end{equation}
and neglecting the $\ep$-dependence of the pre-exponential factors.
Then the integration over $\ep$ gives a $\delta$-function,
$\delta(t_{01}+t_{12}-t_{21'}-t_{1'0})$,
where $t_{ij}$ is the time of travel from the point~$i$ to the
point~$j$ according to the classical equations of motion. This
$\delta$-function simply expresses the fact that the electron and the
hole have to travel for the same amount of time before the radiative
recombination.

\begin{figure}
\includegraphics[width=0.8\linewidth,angle=0,clip]{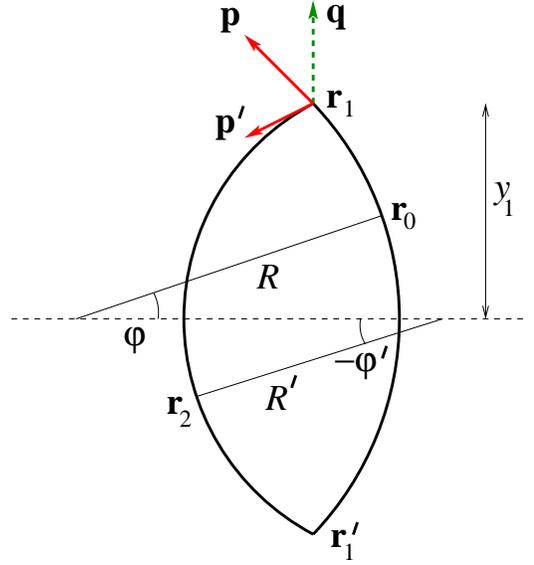}
\caption{\label{fig:stationary}
(Color online) Classical trajectories determining the stationary manifold.}
\end{figure}

Let us assume the phonon momentum $\vec{q}$ to be along the $y$~axis.
Out of eight spatial integration variables two correspond to translations
of the trajectory as a whole. They contribute to normalization, but as we
are not interested here in the overall prefactor in $\mathcal{M}(\vec{q})$,
they can be discarded. In the remaining 6-dimensional space we introduce
three coordinates $y_1,\varphi,\varphi'$ which parametrize the stationary
manifold, and three deviations $\delta{x}_1,\delta{R},\delta{R}'$. The
integration variables are parametrized as (see Fig.~\ref{fig:stationary}):
\begin{eqnarray}
\vec{r}_0&=&\left(R\cos\varphi-\sqrt{R^2-y_1^2},R\sin\varphi\right)\nonumber\\
&&{}+(\delta{R}\cos\varphi,\delta{R}\sin\varphi),\\
\vec{r}_1&=&(0,y_1)+(\delta{x}_1,0),\\
\vec{r}_1'&=&(0,-y_1)+(-\delta{x}_1,0),\\
\vec{r}_0&=&\left(-R'\cos\varphi'-\sqrt{(R')^2-y_1^2},R'\sin\varphi'\right)\nonumber\\
&&{}+(-\delta{R}'\cos\varphi',\delta{R}'\sin\varphi'),
\end{eqnarray}
where
$R=p/(eB/c)$, $R'=p'/(eB/c)$, $p=\win/2v$, $p'=(\win-\omega_\vec{q})/2v$.
If we introduce $\varphi_1=\arcsin(y_1/R)$, $\varphi_1'=\arcsin(y_1/R')$,
then momentum conservation reads as
\begin{equation}
p\sin\varphi_1=p'\sin\varphi_1',\quad
p\cos\varphi_1+p'\cos\varphi_1'=q\,.
\end{equation}
This ensures that the expansion of Eq.~(\ref{exponent=}) in the
deviations does not have linear terms.

Now we have to expand Eq.~(\ref{exponent=}) to the second order in
$\delta{x}_1,\delta{R},\delta{R}'$. If we denote
\begin{equation}
\vec{s}\equiv\frac{\vec{r}-\vec{r}'}{|\vec{r}-\vec{r}'|},
\end{equation}
the second derivatives of the action can be written as
\begin{eqnarray}
&&\frac{\partial{S}}{\partial{x}_i\partial{x}_j}
=\frac{\partial{S}}{\partial{x}_i'\partial{x}_j'}=\nonumber\\
&&=\frac{eB}{2c}\left[-s_is_j\tan\frac{\vartheta_+}2
+(\delta_{ij}-s_is_j)\cot\frac{\vartheta_+}2\right],\\
&&\frac{\partial{S}}{\partial{x}_i\partial{x}_j'}=
\frac{eB}{2c}\times\nonumber\\
&&\qquad\times\left[s_is_j\tan\frac{\vartheta_+}2
-(\delta_{ij}-s_is_j)\cot\frac{\vartheta_+}2-e_{ijz}\right],\nonumber\\
\end{eqnarray}
where $e_{xyz}=-e_{yxz}=1$, $e_{xxz}=e_{yyz}=0$.
As a result, for the quadratic part of the action $\Delta{S}$ we have
\begin{eqnarray}
\Delta{S}&=&-\frac{eB}{2c}\,
\frac{\sin{2}\varphi_1(\delta{R}-\delta{x}_1\cot\varphi_1\sin\varphi)^2}%
{\sin(\varphi_1-\varphi)\sin(\varphi_1+\varphi)}\nonumber\\
&&{}-\frac{eB}{2c}\,
\frac{\sin{2}\varphi_1'(\delta{R}'-\delta{x}_1'\cot\varphi_1'\sin\varphi')^2}%
{\sin(\varphi_1'-\varphi')\sin(\varphi_1'+\varphi')}\nonumber\\
&&{}-\frac{eB}{c}\,(\cot\varphi_1+\cot\varphi_1')(\delta{x}_1)^2.
\end{eqnarray}
After integration over $\delta{R},\delta{R}',\delta{x}_1$, performed
in the Gaussian integration, we are left with three spatial integration
variables $\varphi,\varphi',y_1$, and the energy variable. As mentioned
above, the latter gives the temporal $\delta$-funtion,
\begin{equation}
\delta(t_{01}+t_{12}-t_{21'}-t_{1'0})=\frac{v}2\,\delta(R\varphi+R'\varphi')\,,
\end{equation}
which also lifts the integration over $\varphi'$.
Let us take the limit $\omega_\vec{q}\ll\win$, and neglect the difference
between $R$~and~$R'$ and between $\varphi_1$ and $\varphi_1'$. Choosing
$\varphi_1$ as the integration variable instead of $y_1$, we obtain
\begin{eqnarray}
\mathcal{M}(q)&\propto&\int\limits_0^{\pi/2}R\cos\varphi_1\,d\varphi_1
\int\limits_{-\varphi_1}^{\varphi_1}d\varphi\,
\frac{\sqrt{R\tan\varphi_1}}{\sin{2}\varphi_1}\,\times\nonumber\\
&&{}\times
e^{2ipR(\varphi_1+\sin\varphi_1\cos\varphi_1)-2iqR\sin\varphi_1-4\gamma{R}\varphi_1/v}\times
\nonumber\\
&&{}\times
\cos^2\varphi_1\cos(\varphi_\mathrm{in}-\varphi)\cos(\varphi_\mathrm{out}-\varphi).
\end{eqnarray}
The first integration is simply $R\cos\varphi_1\,d\varphi_1=dy_1$,
the factor $\sqrt{R\tan\varphi_1}/\sin{2}\varphi_1$ comes from
the gaussian integration, and the cosines give the angular dependence
of the electron-photon [$\cos(\varphi_\mathrm{in,out}-\varphi)$] and
electron-phonon ($\cos\varphi_1$) matrix elements. Integration
over $\varphi$ gives
\begin{eqnarray}
\mathcal{M}(q)&\propto&\int\limits_0^{\pi/2}R^{3/2}\,d\varphi_1\,
\sqrt{\sin\varphi_1\cos^3\varphi_1}\times\nonumber\\
&&{}\times
e^{2ipR(\varphi_1+\sin\varphi_1\cos\varphi_1)-2iqR\sin\varphi_1-4\gamma{R}\varphi_1/v}
\times\nonumber\\
&&{}\times
\left[\frac{\cos(\varphi_\mathrm{in}-\varphi_\mathrm{out})}{(\sin\varphi_1)/\varphi_1}
+\cos\varphi_1\cos(\varphi_\mathrm{in}+\varphi_\mathrm{out})\right].
\nonumber\\
\end{eqnarray}
Taking advantage of the limit $\gamma{R}/v\gg{1}$, we can focus on short arcs,
so that $\varphi_1\ll{1}$ and $|2p-q|\ll{p}$. Thus, we expand the exponent to
$\varphi_1^3$, while in the prefactor we keep the leading term at $\varphi_1\to{0}$.
In the experiment described in the paper, the excitation and detection are
unpolarized, so we simply omit the the expression in the square brackets,
which describes the polarization dependence.
Denoting $R\varphi_1=z$, we arrive at
\begin{equation}
\mathcal{M}(q)\propto\int\limits_0^\infty{d}z\,\sqrt{z}\,
e^{-[i(q-2p)+2\gamma/v]z-i[p/(12R^2)]z^3},
\end{equation}
which is Eq.~(1) of the main text. The integral is calculated using the
relations\cite{Reid,AiryBook}
\begin{eqnarray}
&&\frac{1}{2\pi^{3/2}}\int\limits_0^\infty\frac{dt}{\sqrt{t}}\,
\cos\!\left(xt+\frac{t^3}{12}+\frac\pi{4}\right)=\mathrm{Ai}^2(x),\\
&&\frac{1}{2\pi^{3/2}}\int\limits_0^\infty\frac{dt}{\sqrt{t}}\,
\sin\!\left(xt+\frac{t^3}{12}+\frac\pi{4}\right)=\mathrm{Ai}(x)\,\mathrm{Bi}(x).\qquad
\end{eqnarray}

\end{document}